\documentclass{article}

\usepackage{cite}
\usepackage{amsmath,amssymb,amsfonts}
\usepackage{algorithmic}
\usepackage{graphicx}
\usepackage{textcomp}
\usepackage{xcolor}
\usepackage{booktabs}
\usepackage[a4paper, total={6in, 9in}]{geometry}

\begin{document}
\title{A Near-Infrared Enhanced Silicon Single-Photon Avalanche Diode with a Spherically Uniform Electric Field Peak}
\author{
	Edward~Van~Sieleghem$^{1,2}$, Andreas~S\"uss$^{3}$, Pierre~Boulenc$^{4}$, Jiwon~Lee$^{1}$,\\
	Gauri~Karve$^{1}$, Koen~De~Munck$^{1}$, Celso~Cavaco$^{1}$, Chris~Van~Hoof$^{1,2}$\\
	\small$^1$Imec, 3001 Leuven, Belgium\\
	\small$^2$Department of Electrical Engineering, Katholieke Universiteit Leuven, 3001 Leuven, Belgium\\
	\small$^3$OmniVision Technologies, Santa Clara, CA 95054 USA\\
	\small$^4$Gpixel NV, 2018 Antwerpen, Belgium\\
}
\date{}
\maketitle


\textbf{\textcopyright 2021 IEEE.  Personal use of this material is permitted.  Permission from IEEE must be obtained for all other uses, in any current or future media, including reprinting/republishing this material for advertising or promotional purposes, creating new collective works, for resale or redistribution to servers or lists, or reuse of any copyrighted component of this work in other works.\newline}

\begin{abstract}
A near-infrared (NIR) enhanced silicon single-photon avalanche diode (SPAD) fabricated in a customized 0.13~µm CMOS technology is presented. The SPAD has a depleted absorption volume of approximately 15~µm $\times$ 15~µm $\times$ 18~µm. Electrons generated in the absorption region are efficiently transported by drift to a central active avalanche region with a diameter of 2~µm. At the operating voltage, the active region contains a spherically uniform field peak, enabling the multiplication of electrons originating from all corners of the device. The advantages of the SPAD architecture include high NIR photon detection efficiency (PDE), drift-based transport, low afterpulsing, and compatibility with an integrated CMOS readout. A front-side illuminated device is fabricated and characterized. The SPAD has a PDE of 13\% at wavelength 905~nm, an afterpulsing probability $<$~0.1\% for a dead time of 13~ns, and a median dark count rate (DCR) of 840~Hz at room temperature. The device shows promising performance for time-of-flight applications that benefit from uniform NIR-sensitive SPAD arrays.\\
\end{abstract}


\textbf{Funding}: This work was supported in part by Research Foundation -- Flanders (FWO) SB PhD fellowship under Grant 11D3321N.\\

\textbf{Corresponding author}: Edward.Vansieleghem@imec.be

\section{Introduction}
Single-photon avalanche diodes (SPADs) can resolve individual photons with high temporal accuracy. In recent years, various imaging systems exploiting the time-resolving features of SPADs have been demonstrated with increasing spatial resolution. Prominent application domains of SPADs include time-of-flight (ToF) imaging \cite{morimoto2020megapixel,henderson20195,hirose2018250} and biophotonics \cite{bruschini2019single,ulku2018512,gyongy2017256}. In particular, the use of ToF SPAD imagers for range-finding has gained increased interest from the automotive and mobile industries \cite{yole2020lidar, ito2020sonynirspad}. The application benefits from the integration of SPADs into uniform arrays. Infrared illumination is preferred for the reduction of solar background noise and for complying with eye-safety regulations \cite{beer2018background}.

Silicon manufacturing technologies provide a well-established and inexpensive platform for the integration of SPADs. A disadvantage of silicon is the low absorption coefficient in the near-infrared (NIR) spectrum. To optimize the photon detection efficiency (PDE), red and NIR-enhanced silicon SPADs incorporate thick absorption regions using standard \cite{ito2020sonynirspad, lindner2017high, sanzaro2017single, vornicu2019low, abbas2016backside, takai2016single, veerappan2015low, pavia20151, webster2012single, webster2012high} or customized \cite{gulinatti2021custom, lasercomp2020countt, excel2020} technologies. Carriers generated in the absorption volume move towards an electric field peak where they can trigger avalanche breakdown.

Depleting a thick absorption volume in a SPAD potentially results in a high operating voltage, a high excess bias, and sensitivity to fabrication process nonuniformity \cite{gulinatti2021custom}. Such depleted SPAD may require a large guard ring or suffer from junction-edge effects, making integration into dense arrays difficult \cite{morimoto2020high,acerbi2018silicon}. Alternatively, if the absorption volume is not depleted, charge diffusion degrades the timing performance.

In this work, we present a front-side illuminated (FSI) NIR-enhanced silicon SPAD with a depleted absorption volume and an electric field peak enforced by field-line crowding \cite{vansi2019nirspad}. The depleted SPAD can be integrated into arrays with high PDE and low sensitivity to process nonuniformity.

\section{SPAD device}
\begin{figure}[!t]
	\centerline{\includegraphics[width=3.5in]{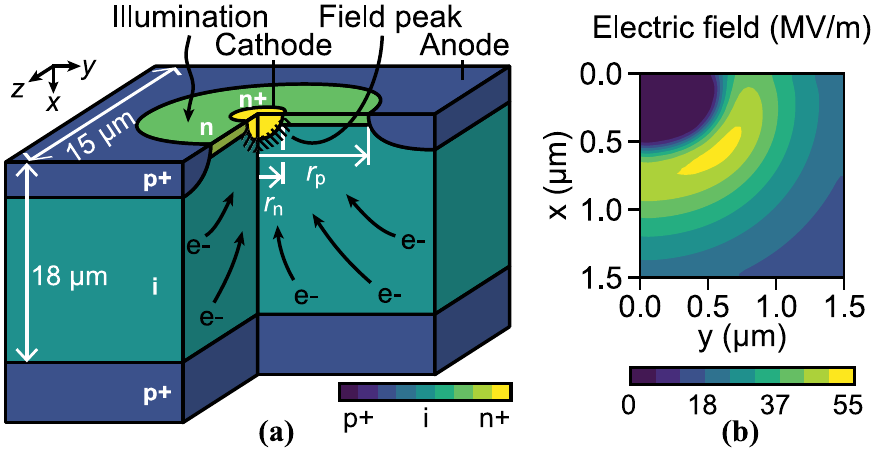}}
	\caption{(a) Doping profile schematic of the SPAD. Electrons (e-) move towards the field peak (dashed area) by drift. (b) Numerical simulation of the electric field near the cathode for $V_{\mathrm{e}}=3.5$~V. The top interface and centerline coincide with $x=0$~µm and $y=0$~µm, respectively.}
	\label{fig:device}
\end{figure}

The doping profile of the NIR-enhanced SPAD is presented in Fig.~\ref{fig:device}(a). The SPAD is formed in an 18~µm-thick intrinsic epitaxial (epi) layer on top of a p+ substrate by using a customized 0.13~µm silicon CMOS technology. The cathode consists of an n+ region in the shape of a half-sphere with radius $r_{\mathrm{n}}=0.6$~µm. The anode is formed by a p+ region concentrically enclosing the cathode with an inner radius $r_{\mathrm{p}}=5$~µm. A shallow n-type implantation is present between the cathode and the anode. The device has a pitch of 15~µm.

The SPAD operates at a reverse bias equal to the sum of the excess bias $V_{\mathrm{e}}$ and the breakdown voltage $V_{\mathrm{bd}}$. At the operating voltage, the epi layer and the shallow n-type region are fully depleted. As a result, the SPAD contains a depleted absorption volume of nearly $15$~µm $\times$ 15~µm $\times$ 18~µm in which an electric field is present. Electrons generated in the absorption region move efficiently towards the cathode by drift, without requiring additional assistance \cite{jeganna2020currentassisted}. The significant thickness of the absorption volume and the drift-based transport enhance the NIR sensitivity and timing performance of the detector, respectively. The presented FSI SPAD is not isolated from the substrate, and diffusion from the substrate affects the timing performance. Besides, the device has no physical isolation to prevent crosstalk.

A spherically uniform electric field peak is present near the cathode, as visualized in Fig.~\ref{fig:device}(b) for $V_\mathrm{e}=3.5$~V. The active high-field region has a diameter of $d_\mathrm{act}=2$~µm. Most electrons entering the uniform field peak have a high probability of triggering avalanche breakdown. This feature resembles electrical microlensing and greatly enhances the PDE \cite{morimoto2020lensing,veerappan2013silicon}. Additionally, the small active volume enables a low afterpulsing probability. The field peak is enforced by field-line crowding as opposed to ionized doping atoms. As a result, the device uniformity across a wafer has a low sensitivity to doping nonuniformity. The cathode has the shape of a halve-sphere to achieve the spherical uniformity of the field. The radius $r_{\mathrm{n}}$ of the cathode largely determines $V_{\mathrm{bd}}$ and the magnitude of the field in the epi at $V_{\mathrm{bd}}$. In this work, the radius is $r_{\mathrm{n}}=0.6$~µm, resulting in $V_{\mathrm{bd}}>60$~V and a sufficient field in the 18~µm epi to achieve 300~ps NIR timing resolution. The breakdown voltage is relatively high and can lead to significant power consumption. Reducing $V_{\mathrm{bd}}$ is possible by making $r_{\mathrm{n}}$ smaller at the cost of slower electron transport, which is acceptable if the epi thickness is reduced.

The field peak is locally reduced on the top interface by the shallow implantation. Carriers generated on the interface have a lower probability of multiplying, reducing the dark count rate (DCR). This feature resembles the field redistribution techniques used in power semiconductor devices \cite{resurf2000}. The dose of the shallow implant is high enough to redistribute the field on the interface but low enough to not negatively affect the PDE.

\section{Test system}
\begin{figure}[!t]
	\centerline{\includegraphics[width=3.5in]{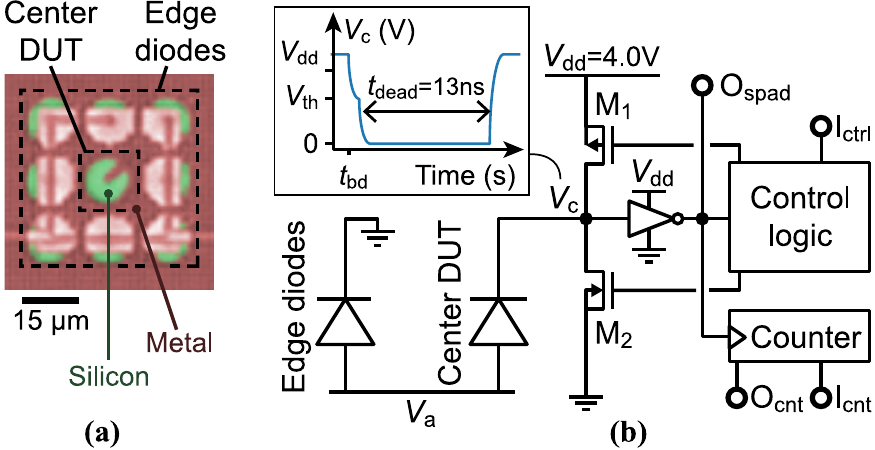}}
	\caption{(a) Micrograph of the diode array. Silicon and metal are artificially colored green and red, respectively. (b) Digital SPAD module. The inset shows the nominal $V_\mathrm{c}$ response for an avalanche event.}
	\label{fig:module}
\end{figure}

The device is integrated into a $3\times 3$ array, as illustrated in Fig.~\ref{fig:module}(a). The cathodes are electrically isolated by potential barriers, whereas the anodes and substrate are electrically connected. The central diode is the device under test (DUT). Two types of test modules are considered. Firstly, wired-out modules are used for current-voltage (IV) characterization. Herein, the DUT cathode is directly connected to a source measurement unit. Secondly, digital SPAD modules are used for avalanche event characterization. Herein, the DUT cathode is connected to an integrated circuit as discussed further. 

Fig.~\ref{fig:module}(b) illustrates the monolithic integrated circuit of a digital SPAD module. The circuit is isolated from the substrate by a buried n-well process. The cathode of the DUT is connected to two transistors and a CMOS inverter. The output of the inverter is connected to a quench/recharge control circuit, a 30-bit asynchronous counter, and a buffered output pad O$_{\mathrm{spad}}$. The control circuit determines whether the cathode voltage $V_{\mathrm{c}}$ is pulled to the supply voltage $V_{\mathrm{dd}}=4.0$~V or ground by turning on transistor M$_1$ or M$_2$, respectively. In the idle state, both M$_1$ and M$_2$ are off, and the relative sizing of the transistors ensures $V_{\mathrm{c}} \approx V_{\mathrm{dd}}$.

The anode voltage $V_{\mathrm{a}}$ is configured such that the reverse bias of the DUT idles above the breakdown voltage. The excess bias is defined as $V_{\mathrm{e}} = V_{\mathrm{a}} + V_{\mathrm{dd}} - V_{\mathrm{bd}}$. During nominal operation, $V_{\mathrm{e}}$ is selected between $V_{\mathrm{dd}}-V_{\mathrm{th}}$ and $V_{\mathrm{dd}}$, with inverter threshold voltage $V_{\mathrm{th}}=V_{\mathrm{dd}}/2$. The inset of Fig.~\ref{fig:module}(b) shows the nominal cathode response for an avalanche event at time $t_{\mathrm{bd}}$. Initially, transistor M$_1$ is off and acts as a quenching load. When an avalanche event triggers a breakdown current, $V_{\mathrm{c}}$ reduces below $V_{\mathrm{th}}$ due to passive quenching. The output of the inverter flips, and transistor M$_2$ is turned on by the control circuit. Consequently, the cathode is pulled to the ground, quenching the avalanche current. After a fixed dead time of $t_{\mathrm{dead}}=13$~ns, M$_1$ is briefly turned on and M$_2$ is turned off by the control circuit. These actions pull the cathode back to $V_{\mathrm{dd}}$. Each rising edge of the inverter output increments the counter value at O$_\mathrm{cnt}$ and produces a signal at O$_\mathrm{spad}$. The counter is only incremented if $V_{\mathrm{e}}>V_{\mathrm{dd}}/2$. This boundary condition is used for calibrating $V_{\mathrm{e}}$. Throughout this work, the edge diodes are biased below $V_\mathrm{dd}$. Crosstalk is prevented since these diodes are not in breakdown and contain a similar field as the DUT.

\section{Results}
\begin{figure}[!t]
	\centerline{\includegraphics[width=3.5in]{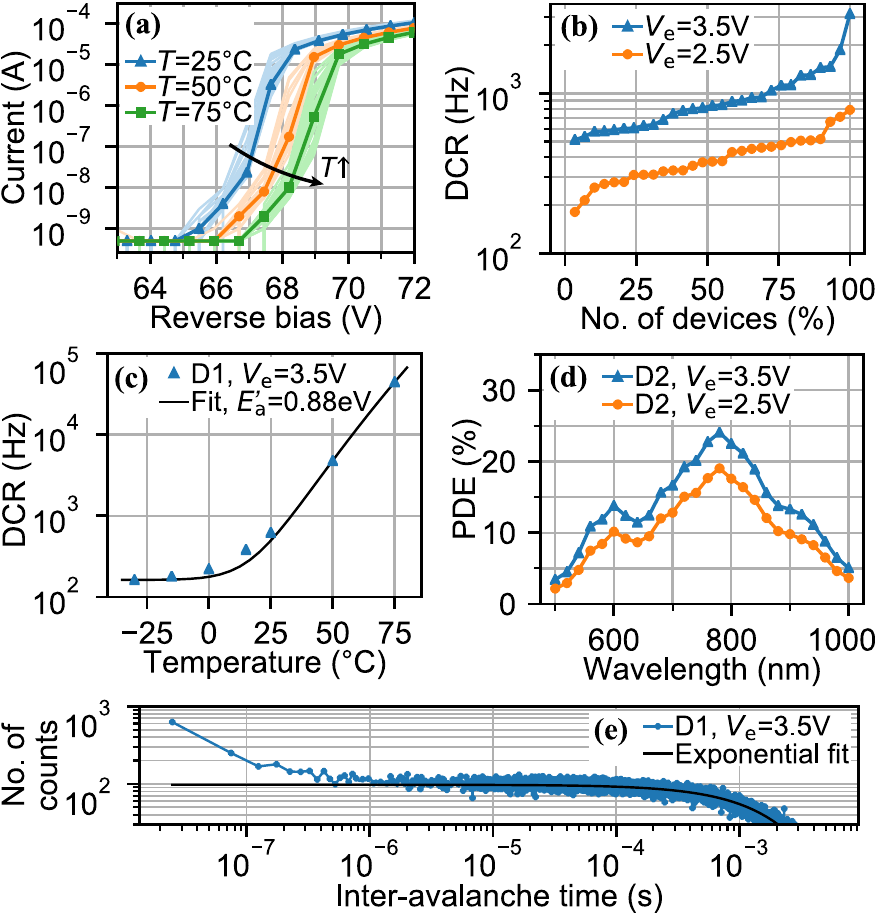}}
	\caption{Characterization results of the SPAD. (a) IV characteristics in the dark versus $T$ with a resolution of 1~nA. The median IV is highlighted. (b) DCR distribution for $T=25$~°C. (c) DCR versus $T$ for device D1. (d) PDE versus wavelength for device D2 and $T=25$~°C. (e) Inter-avalanche time histogram with time-bin 50~ns obtained from $3.5\times10^6$~events in device D1.}
	\label{fig:result_1}
\end{figure}

The quasi-static IV characteristics of 29 devices on a wafer have been measured. Fig.~\ref{fig:result_1}(a) presents the median dark current as a function of the reverse bias and temperature $T$. The median breakdown voltage is 67.25~V for $T=25$~°C with a temperature coefficient of 34~mV/K. The breakdown voltages of all devices vary by less than 1\% from the median. The median dark current activation energy is $E_\mathrm{a}=0.67$~eV near $V_\mathrm{bd}$ and $T=25$~°C, which is consistent with Shockley-Read-Hall generation and diffusion. The median reverse resistance above breakdown equals $R_\mathrm{r}=44~\mathrm{k\Omega}$ for $T=25$~°C. The resistance is higher than typical SPADs, and likely results from space-charge and substrate resistances. Numerical simulations predict a junction capacitance $<0.5$~fF, enabling a moderate overall RC-delay constant for quenching.

Fig.~\ref{fig:result_1}(b) shows the cumulative DCR distribution of 29 devices for $T=25$~°C. The median DCR is 840~Hz for $V_{\mathrm{e}}=3.5$~V. Devices D1 and D2 with near-median DCR are characterized further. Fig.~\ref{fig:result_1}(c) shows the DCR temperature dependence of device D1 for $V_{\mathrm{e}}=3.5$~V. The data are fitted by the sum of two terms. The first term is constant and equal to $162$~Hz, which is consistent with trap-assisted tunneling. The second term is exponential and given by $A\times\mathrm{exp}(-E'_\mathrm{a}/(k_\mathrm{b}T))$ with activation energy $E'_\mathrm{a}=0.88$~eV, Boltzmann constant $k_\mathrm{b}$, and fitting parameter $A$. This term is consistent with SRH generation and diffusion, and it is dominant for $T>0$~°C. The inequality $E_\mathrm{a}<E'_\mathrm{a}$ indicates that the dark carriers generated by SRH on the interface have a lower probability of contributing to the DCR due to the locally redistributed field. The DCR related to diffusion can be reduced by optimizing the doping profile and by removing the substrate.

Fig.~\ref{fig:result_1}(d) illustrates the PDE obtained by illuminating device D2 with a homogeneous monochromatic source. The PDE at wavelength $\lambda=905$~nm is 13\% for $V_{\mathrm{e}}=3.5$~V. The large depletion region and uniform field peak enable the high NIR sensitivity. Dummy metal layers, serving no functional purpose in the FSI device, limit the PDE by reflecting approximately half of the incident light on the DUT.

The afterpulsing probability $P_\mathrm{ap}$ of device D1 is calculated based on the time difference between consecutive avalanche events \cite{fishburn2012fundamentals}. Fig.~\ref{fig:result_1}(e) shows an inter-avalanche time histogram for $V_{\mathrm{e}}=3.5$~V, $T=25$~°C, and $t_{\mathrm{dead}}=13$~ns. The histogram is fitted by an exponential function corresponding to Poisson distributed events. The area above the fit corresponds to afterpulsing with $P_\mathrm{ap}<0.1\%$. Very few carriers are trapped in the small active region during avalanche events.

The charge transport in the absorption volume is based on drift. However, electrons generated in the substrate can also diffuse towards the cathode. Numerical simulations predict a NIR full width at half maximum (FWHM) timing resolution $<300$~ps and a diffusion tail $<2$~ns. The resistance $R_\mathrm{r}$ may negatively affect the timing performance. Besides, due to the lack of physical isolation between neighboring devices, crosstalk is significant. However, the edge diodes and DUT contain similar electric fields during operation, and there is no indication that the presented results are biased by crosstalk.

\section{Conclusion}
Table. \ref{tab:comparison} provides a comparison of NIR-sensitive silicon SPADs. Despite the degradation of the PDE by metal reflections, the presented FSI device achieves a competitive NIR PDE. Additionally, $P_\mathrm{ap}$ exceeds the state-of-the-art \cite{ito2020sonynirspad} due to the small active region. Unlike typical SPADs with thick depleted absorption volumes \cite{gulinatti2021custom}, the device is compatible with a standard CMOS readout (enabled by $V_\mathrm{e}=3.5$~V), and the pitch is not limited by a guard ring. The lack of isolation negatively affects the NIR timing resolution and crosstalk of the FSI SPAD, as is the case for other nonisolated SPADs \cite{takai2016single, webster2012single}. However, since the absorption volume is depleted, back-side illumination (BSI) and deep-trench isolation (as in \cite{ito2020sonynirspad}) can greatly improve isolation while enhancing the PDE and timing resolution. Additionally, $V_\mathrm{bd}<40$~V and a lower $R_\mathrm{r}$ can be obtained when reducing the epi thickness and optimizing the doping. All considered, the presented device shows promise for ToF applications. The fabrication of a BSI device with a monolithic readout circuit is ongoing. The timing jitter and crosstalk will also be characterized in the future.

\begin{table}
	\centering
	\caption{Comparison of NIR-sensitive silicon SPADs.}
	\begin{tabular}{c|ccccc}
		\hline
		& \textbf{This} & \cite{gulinatti2021custom} & \cite{ito2020sonynirspad} & \cite{takai2016single} & \cite{webster2012single} \\
		& \textbf{work}& 2021 & 2020 & 2016 & 2012 \\
		\hline
		\hline
		Technology & 130nm & custom & 90nm & 180nm & 90nm  \\
		\hline
		$V_\mathrm{bd}$+$V_\mathrm{e}$ (V) & 67.3+3.5 & 30+20 & 20+3 & 20.5+5 & 15+2.4 \\
		\hline
		DCR (Hz) & 840& 3300 & 3 & 125 & 1000 \\
		\hline
		PDE@905nm (\%) & 13 & 20‡ & 20.5 & 4.3 & 13‡ \\
		\hline
		Pitch/$d_\mathrm{act}$ (µm) & 15/2 & 250/50 & 10/5 & 25/14 & -/6.4 \\
		\hline
		$P_\mathrm{ap}$ (\%) & $<0.1$ & 2 & $0.1$ & - & 0.4 \\
		\hline
		Jitter FWHM & 300* & 95 & 173 & - & 51 \\
		@$\lambda$ (ps@nm) & @905 & @820 & - & - & @470† \\
		\hline
	\end{tabular}\\
	*Predicted by Monte-Carlo simulations,\\†Not representative for NIR applications,\\‡Assuming fill-factor = 1.
	\label{tab:comparison}
\end{table}

\end{document}